\documentclass[10pt,a4paper]{article}

\usepackage{a4,amsmath,bbm,times,graphicx}
\newcommand{\one}{\mbox{$1 \hspace{-1.0mm}  {\bf l}$}}  

\begin{document}

\bigskip\bigskip\bigskip

\noindent 
{\LARGE{\bf One-way Quantum Computation}}

\bigskip\bigskip

\noindent
{{Dan  Browne$^a$ and Hans  Briegel$^b$}}

\bigskip

\noindent
{$^a$Departments of Materials and Physics, Oxford University, United Kingdom.\\
$^b$Institute for Theoretical Physics, University of Innsbruck  and
Institute for Quantum Optics \& Quantum Information (IQOQI)
of the Austrian Academy of Sciences, Austria.}

\bigskip\bigskip
\bigskip\bigskip
\hrule

\bigskip

\def\dbra#1{{\langle{#1}|}}
\def\dket#1{{|{#1}\rangle}}
\newcommand{\dbraket}[2]{\langle #1|#2\rangle}




\section{Introduction}

The  circuit model of quantum computation \cite{brownebriegel_deutschqcn,brownebriegel_barenco,brownebriegel_nielsenchuang} has been a powerful tool for the development of quantum computation, acting both as a framework for theoretical investigations and as a guide for experiment. In the circuit model (also called the network model), unitary operations are represented by a network of elementary quantum gates such as the CNOT gate and single-qubit rotations. 
Many proposals for the implementation of quantum  computation are designed around this model, including physical prescriptions for implementing the elementary gates. By formulating quantum computation in a different way, one can gain both a new framework for experiments and new theoretical insights. \emph{One-way quantum computation} \cite{brownebriegel_clusterqc1} has achieved both of these.

Measurements on entangled states play a key role in many quantum information protocols, such as quantum  teleportation and entanglement-based quantum  key distribution. In these applications an entangled state is required, which must be generated beforehand. Then, during the protocol, measurements are made which convert the quantum correlations into, for example, a secret key. To repeat the protocol a fresh entangled state must be prepared. In this sense, the entangled state, or the quantum correlations embodied by the state, can be considered a resource which is ``used up'' in the protocol.

In one-way quantum computation, the quantum correlations  in an entangled state  called a \emph{cluster state} \cite{brownebriegel_briegelcluster} or \emph{graph state} \cite{brownebriegel_hein} are exploited to allow universal quantum computation through single-qubit measurements alone. The quantum algorithm is specified in the choice of bases for these measurements and  the ``structure'' of the entanglement (as explained below) of the resource state. The name ``one-way''  reflects the resource nature of the graph state. The state can be used only once, and (irreversible) projective measurements drive the computation forward, in contrast to the reversibility of every gate in the standard network model.

In this chapter, we will provide an introduction to one-way quantum computation, and several of the techniques one can use to describe it. In this section we will introduce graph and cluster states and develop a notation for general single-qubit measurements. In section~\ref{brownebriegel_simple} we will introduce the key concepts of one-way quantum computation with some simple examples. After this, in section~\ref{brownebriegel_beyond}, we shall investigate how one-way quantum computation can be described without using the quantum circuit model. To this end, we shall introduce a number of important tools including the stabilizer formalism, the logical Heisenberg picture and  a representation of unitary operations especially well suited to the one-way quantum computation model. In  section~\ref{brownebriegel_implementations}, we will briefly describe a number of proposals for implementing one-way quantum computation in the laboratory. 
In section~\ref{recentdev} we will conclude with a brief survey of some recent research developments in measurement-based quantum computation.

A different perspective of one-way quantum computation and measurement-based computation in general can be found in these recent reviews \cite{brownebriegel_nielsenreview,brownebriegel_jozsareview}. A comprehensive tutorial and review on the properties of graph states can be found in \cite{brownebriegel_graphtutorial}.

\subsection{Cluster states and graph states}\index{Cluster state}\index{Graph state}

Cluster states and graph states can be defined constructively in the following way \cite{brownebriegel_briegelcluster,brownebriegel_graphtutorial}. With each state, we associate a  graph, a set of vertices and edges connecting vertex pairs. Each vertex on the graph corresponds to 
 a qubit. The corresponding ``graph state'' may be generated by preparing every qubit in the state $\dket{+}=(1/\sqrt{2})(\dket{0}+\dket{1})$ and   applying a controlled $\sigma_z$ (CZ) operation $\dket{0}\dbra{0}\otimes\one+\dket{1}\dbra{1}\otimes\sigma_z$ on  every pair of qubits whose vertices are connected by a graph edge. Cluster states are a sub-class of graph states, whose underlying graph is an $n-$dimensional square grid.  The extra flexibility in the entanglement structure of graph states means that they often require far fewer qubits to implement the same one-way quantum computation. However, there are a number of physical implementations where the regular layout of cluster states means that they can be generated very efficiently (see section~\ref{brownebriegel_implementations}).

\begin{figure}
\begin{center}
\includegraphics[width=12cm]{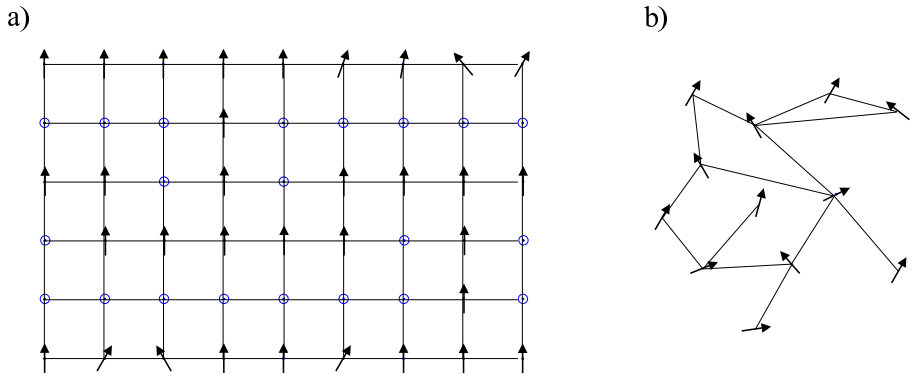}
\end{center}
\caption{\label{brownebriegel_clustervsgraph} One way quantum computation consists of single-qubit measurements in certain bases and in a certain order on an entangled resource state. Cluster states have a square lattice structure (a) while the freedom of choosing specific general  graph states such as illustrated in (b) can reduce the number of qubits needed for a given computation significantly.}
\end{figure}

\subsection{Single-qubit measurements and rotations}\label{brownebriegel_singlequbmeas}

\begin{figure}
\begin{center}
\includegraphics{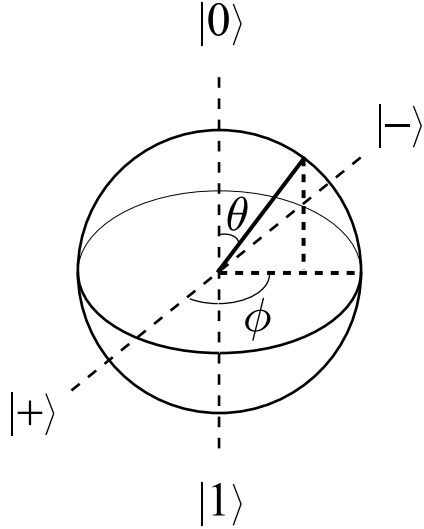}
\end{center}
\caption{\label{brownebriegel_blochmeas} Single-qubit projective measurements will be represented by the pair of angles $(\theta,\phi)$ of the co-latitude $\theta$ and longitude $\phi$  of their $+1$ eigenstate on the Bloch sphere. This corresponds to a measurement of the observable $U_z(\phi+\pi/2)U_x(\theta)ZU_x(-\theta)U_z(-\phi-\pi/2)$. }
\end{figure}

Single-qubit measurements in a variety of bases play a key role in one-way quantum computation, so here we introduce a convenient and compact way to describe them. Using a Bloch sphere\index{Bloch sphere} picture, every projective single-qubit measurement can be associated with a unit vector  on the sphere, which corresponds to the $+1$ eigenstate of the measurement. We can then parameterize observables by the co-latitude $\theta$ and longitude $\phi$ of this vector  (illustrated in figure~\ref{brownebriegel_blochmeas}). We shall write this compactly as a pair of angles $(\theta,\phi)$. 

Unitary operations corresponding to rotations on the Bloch sphere have the following form. A rotation around the $k$ axis (where $k$ is $x$, $y$, or $z$) by angle $\phi$ can be written\begin{equation}\label{brownebriegel_eqrot}
U_k(\phi)=e^{-\frac{i \phi}{2} \sigma_k}
\end{equation}
For brevity and clarity, we will use the notation $X\equiv\sigma_x$, etc. in the rest of this chapter. We also adopt standard notation for the eigenstates of $Z$ and $X$:
\begin{equation}
\begin{split}
Z&\dket{0}=\dket{0}\\
-Z&\dket{1}=\dket{1}\\
X&\dket{+}=\dket{+}\equiv\tfrac{1}{\sqrt{2}}(\dket{0}+\dket{1})\\
-X&\dket{-}=\dket{-}\equiv\tfrac{1}{\sqrt{2}}(\dket{0}-\dket{1})
\end{split}
\end{equation}

A measurement with angles $(\theta,\phi)$  corresponds to a measurement of the observable $U_z(\phi+\pi/2)U_x(\theta)ZU_x(-\theta)U_z(-\phi-\pi/2)$. One way of implementing such a measurement is to apply the single-qubit unitary $U_x(-\theta)U_z(-\phi-\pi/2)$  to the qubit before measuring it in the computational basis.

\section{Simple examples}\label{brownebriegel_simple}

Many of the features of  one-way quantum computation can be illustrated  in a simple two-qubit example.
Consider the following simple protocol; a qubit is prepared in  an unknown state $\dket{\psi}=\alpha\dket{0}+\beta\dket{1}$.  A second qubit is prepared in the state $\dket{+}=\tfrac{1}{\sqrt{2}}(\dket{0}+\dket{1})$. A CZ operation is applied on the two qubits.The state of the qubits is then
\begin{equation}
\frac{1}{\sqrt{2}}\left(\alpha\dket{0}\dket{{+}}+\beta\dket{1}\dket{{-}}\right)\ .
\end{equation}

The 
 first qubit is now measured in the basis $\{(1/\sqrt{2})(\dket{0}\pm e^{i \phi}\dket{1})\}$, where $\phi$ is a real parameter.  Using the notation introduced in section~\ref{brownebriegel_singlequbmeas} this measurement is denoted $(\pi/2,\phi)$. This corresponds, in the Bloch sphere picture, to a unit vector in the $x$-$y$ plane at angle $\phi$ to the $x$ axis.
There are two possible outcomes to the measurement, which occur with equal probability. If the measurement returns the $+1$ eigenvalue, the second qubit will be projected into the state
\begin{equation}
\alpha\dket{+}+ e^{i\phi}\beta\dket{-}\ .
\end{equation}
If the $-1$ eigenvalue is found the state of qubit two becomes
\begin{equation}
\alpha\dket{+}- e^{i\phi}\beta\dket{-}\ .
\end{equation}
We can represent both possibilities in a compact way if we introduce the binary digit $m\in\{0,1\}$ to represent a measurement outcome of $(-1)^m$. The state of qubit two can then be written, up to a global phase, 
\begin{equation}
X^m H U_z(\phi)\dket{\psi}\ .
\end{equation}

We see that the unknown input state which was prepared on the first  qubit  has been  transferred to qubit two without any loss of coherence. In addition to this it has undergone a  unitary transformation: $X^m H U_z(\phi)$. Notice that the angle of the rotation  $U_z(\phi)$ is set in the choice of measurement basis. The unitary transformation  $H U_z(\phi)$ is accompanied by an additional Pauli transformation ($X$) when the measurement outcome is $-1$. This is a typical feature of one-way quantum computation; due to the randomness of the measurement outcomes, any desired unitary can be implemented only up to random but known Pauli transformations. Since these Pauli operators are an undesired by-product of implementing the unitary in the one-way model, we call them ``by-product operators'' \cite{brownebriegel_clusterqc1,brownebriegel_clusterqc2}. As we shall see below, these extra Pauli operations can be accounted for by altering the basis of later measurements, making the scheme deterministic but introducing an unavoidable time-ordering.

\begin{figure}
\begin{center}
\includegraphics[scale=0.2]{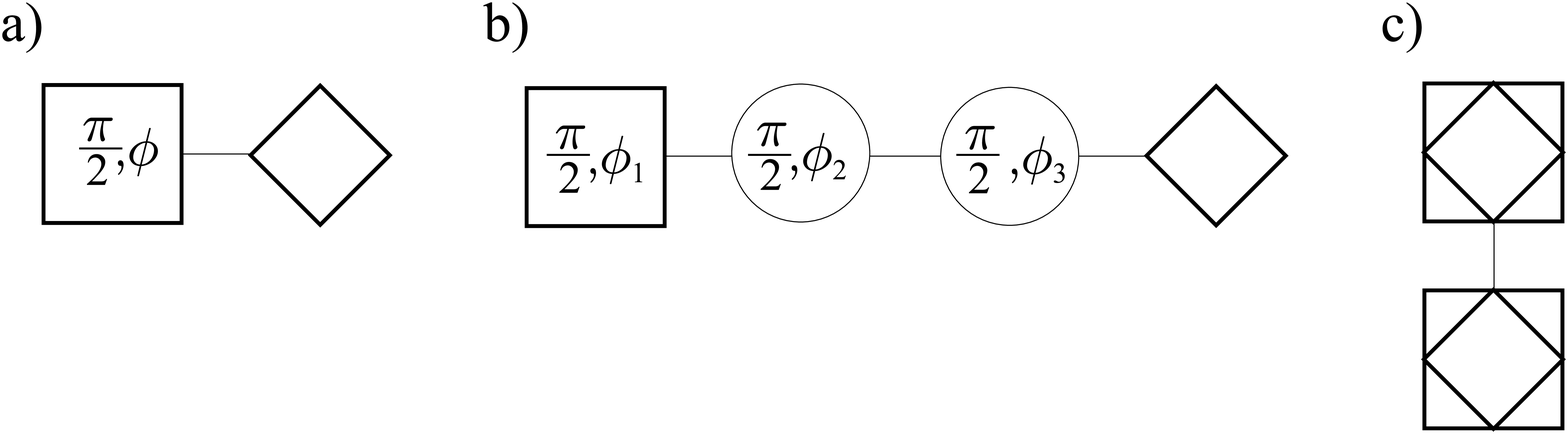}
\end{center}
\caption{\label{brownebriegel_simplepatterns} The one-way graph and measurement patterns for a) the single-qubit operation $H U_z(\phi)$ and b) an arbitrary single-qubit operation, $U_z(\gamma)U_x(\beta)U_z(\alpha)$, when the measurement angles are set $\phi_1=\alpha$, $\phi_2=(-1)^{m_1}\beta$ and $\phi_3=(-1)^{m_2}\gamma$, and $m_a$ is the binary measurement outcome of the measurement on qubit $a$. Note that this imposes an ordering in the measurements of this pattern.
 This second pattern is made by composing three copies of the pattern (a) with differing measurement angles as described in the text. Pattern (c) implements a CZ operation. Input and output qubits coincide for this pattern. }
\end{figure}

In figure~\ref{brownebriegel_simplepatterns} (a), this protocol is represented  using a graphical notation that we will use throughout this chapter. The input qubit is represented by a square, and the output qubit by a lozenge, a smaller square tilted at $45^\circ$. The CZ operation applied to the two qubits is represented by a line between them. This is an example of a one-way graph and measurement pattern, or ``one-way pattern'' for short,  a convenient representation which specifies both the entanglement graph for the resource state and the measurements required to  implement a  unitary operation (always up to a known but random Pauli transformation) in the one-way model.  As an alternative to this graphical approach, an algebraic representation of one-way patterns called  the ``measurement calculus'' has been developed recently \cite{brownebriegel_kashefidanos}.

 So far, the protocol described above seems rather different from the description of one-way quantum computation as a series of measurements on a special entangled resource state. We shall  see below how the two pictures are related. First however, we show how one-way patterns may be connected together to perform consecutive operations.

\subsection{Connecting one-way patterns - arbitrary single-qubit operations}

Due to Euler's rotation theorem any single-qubit SU(2) rotation can be decomposed as a product of three rotations $U_z(\gamma)U_x(\beta)U_z(\alpha)$. Thus, by repeating the simple two-qubit protocol three times, any arbitrary single-qubit rotation may be obtained (up to an extra Hadamard, which can  be accounted for). Two one-way patterns are combined as one would expect, the output qubit(s) of one pattern become the input qubit(s) of the next. The main issue in connecting patterns together is to track the effect of the  Pauli by-product operators which have accumulated due to the previous measurements.

Concatenating the two-qubit protocol three times, with different angles $\phi_1$,  $\phi_2$ and  $\phi_3$ gives the one-way pattern illustrated in figure~\ref{brownebriegel_simplepatterns} (b). To see the effect of the by-product operators from each measurement, let us label the binary outcome from each $m_a$. The  unitary implemented by the combined pattern is therefore
\begin{equation}
U=H Z^{m_3} U_z(\phi_3)H Z^{m_2} U_z(\phi_2)H Z^{m_1} U_z(\phi_1).
\end{equation}
Since $HZH=X$ and $H U_z(\phi)H=U_x(\phi)$ this can be rewritten
\begin{equation}
H Z^{m_3} U_z(\phi_3) X^{m_2} U_x(\phi_2) Z^{m_1} U_z(\phi_1).
\end{equation}
We can rewrite this further using the identities $X U_z(\phi)= U_z(-\phi) X$ and $Z U_x(\phi)= U_x(-\phi) Z$,
\begin{equation}
 X^{m_3} Z^{m_2}  X^{m_1} H U_z((-1)^{m_2}\phi_3)  U_x((-1)^{m_1}\phi_2) U_z(\phi_1).
\end{equation}

Now we have split up the operation in the same way as the two-qubit example, a  unitary  plus a known  Pauli correction. In this case, however, this unitary is not deterministic -- the sign of two of the rotations depends on two of the measurements. Nevertheless, if we perform the measurements sequentially and choose  measurement angles $\phi_1=\alpha$, $\phi_2=(-1)^{m_1}\beta$ and   $\phi_3=(-1)^{m_2}\gamma$, we obtain deterministically the desired single-qubit unitary.

The dependency of measurement bases on the outcome of previous measurements is a generic feature of one-way quantum computation, occurring for all but a special class of operations, the Clifford group (described below). This dependency means that there is a minimum number of time-steps in which any one-way quantum computation can be implemented, as discussed further in section~\ref{brownebriegel_beyond}.

The Pauli corrections remaining at the  end of the implemented one-way quantum computation
 are unimportant and never need to be physically applied; they can always be accounted for in the interpretation of the final measurement outcome. For example, if the final state is to be read out in the computational basis any extra  $Z$ operations commute with the measurements and have no effect on their outcome. Any $X$ operations simply flip the measurement result, and thus can be corrected via classical post-processing. 

\subsection{Graph states as a resource}

It is worth discussing how the above description of one-way patterns relates to the description of one-way quantum computation in the introduction, namely as measurements on an entangled resource state. The first observation is that, given a one-way pattern, all of the measurements can be made after all the CZ operations have been implemented. Secondly, quantum algorithms usually begin by initializing qubits to a fiducial starting state. This state is usually $\dket{0}$ on each qubit, but the state $\dket{+}$ would be equally suitable. When the input qubits of a one-way graph measurement pattern are prepared in $\dket{+}$, then the entangled state generated by the CZ gates  is a graph state. Thus the graph state can be considered a resource for this  quantum computation. We shall see in section~\ref{brownebriegel_implementations} that for certain implementations, such as in linear optics, the resource description is especially apt.

\subsection{Two-qubit gates}
So far we have seen  how an arbitrary single-qubit operation could be achieved in one-way quantum computation in a simple linear one-way pattern.  However, for universal quantum computation,   entangling two-qubit gates are necessary. One such gate is a CZ gate. There is a particularly simple way in which the CZ can be implemented within the one-way framework. This is simply to use the CZ represented by a single graph-state edge  to implement the CZ directly. This leads to the one-way pattern illustrated in figure~\ref{brownebriegel_simple} (c). Notice that here the input qubits are also the output qubits. This is indicated by the superimposed squares and lozenges.

\subsection{Cluster-state quantum computing}
In a number of proposed implementations  of one-way quantum computation (see section~\ref{brownebriegel_implementations})  square lattice cluster states can be generated efficiently and arbitrarily connected graph states are hard to make.
The simple method outlined above for the construction of one way patterns will usually lead to graph state layouts which do not have a square lattice structure. Nevertheless, a cluster state on a large enough square lattice of two or more dimensions is still sufficient to implement any unitary \cite{brownebriegel_clusterqc1}.
 A number of measurement patterns for quantum gates designed specifically for two-dimensional square lattice cluster states can be found in  \cite{brownebriegel_clusterqc2}.

\section{Beyond quantum circuit simulation}\label{brownebriegel_beyond}

We have shown that  the one-way quantum computer can implement deterministically a universal set of gates and thus any quantum computation. However, part of the power of  one-way quantum computation derives from the fact that unitary operations can be implemented more compactly than a naive network construction would suggest \cite{brownebriegel_jmopaper}. In fact we shall see in the following sections that other ways of  decomposing  unitary operations   are more natural and useful. The main tool we shall use in our  investigation of these properties is the \emph{stabilizer formalism}.

\subsection{Stabilizer formalism}\index{Stabilizer Formalism}

The stabilizer formalism \cite{brownebriegel_gottthes,brownebriegel_gottheis} is a powerful tool for understanding the properties of graph states and one-way quantum computation. Stabilizer formalism is a framework whereby  states and sub-spaces over multiple qubits are described and characterized in a compact way in terms of  operators under which they are invariant. In standard quantum mechanics one uses  complete sets of commuting observables in a similar fashion, such as in the description of atomic states by ``quantum numbers'' (see e.g. \cite{brownebriegel_cohen}). 

An operator $K$ \emph{stabilizes} a subspace $\mathcal{S}$ when, for all states $\dket{\psi}\in\mathcal{S}$,
\begin{equation}\label{brownebriegel_stabeig}
K\dket{\psi}=\dket{\psi}.
\end{equation}
In other words, $\dket{\psi}$ is an eigenstate of $K$ with eigenvalue $+1$. 

In the stabilizer formalism one focuses on  operators which, in addition to this stabilizing property, are  Hermitian members of the Pauli group, i.e. tensor products of Pauli and identity
operators. The key principle of the stabilizer formalism is to identify a set of such stabilizing operators which uniquely defines a given state or sub-space  - i.e. there is no state outside the sub-space (for a specified system) which the same set of operators also jointly stabilizes. The  sub-spaces (and states) which can be defined uniquely using stabilizing operators from the Pauli group are called \emph{stabilizer sub-spaces} (or stabilizer states).
 
Stabilizer states and sub-spaces occur widely in quantum information science and include Bell states, GHZ states, many error-correcting codes, and, of course,  graph states and cluster states.  
Note that  there are other joint eigenstates of the stabilizing operators with some $-1$ eigenvalues. However, only states with $+1$ eigenvalue are ``stabilized'', by definition.
 This set of operators  then embodies all the properties of the state and  can allow an easier analysis, for example, of how the state transforms under  measurement and unitary evolution. 
Since the product of two stabilizing operators is itself  stabilizing, the set of operators which stabilize a sub-space has a group structure. It is called the \emph{stabilizer group} or simply the \emph{stabilizer} of the sub-space. The
 group can be compactly expressed by identifying a set of generators. For a $k$-qubit  sub-space in an $n$ qubit system, $n-k$ generators are required (see exercise 2).

We do not have enough space here for a detailed introduction to all of the techniques of stabilizer formalism -- excellent introductions can be found in  \cite{brownebriegel_nielsenchuang,brownebriegel_gottthes} -- but instead we will  focus on those which are useful for understanding one-way quantum computation. Most will be stated without proof but can be verified using the  properties of Pauli group operators described in  \cite{brownebriegel_nielsenchuang}.

A simple example of a stabilizer state is the state $\dket{+}$. Its stabilizer group is generated by $X$ alone. The stabilizer  for the tensor product state $\dket{+}^{\otimes n}$ is then generated by $n$ operators $K_a=X_a$ acting on each qubit $a$. From this we can derive the stabilizer generators for graph states. Consider a stabilizer state transformed by the unitary transformation $V$.  The stabilizers of the transformed state are then given by $VK_aV^\dag$. Since the CZ gate transforms $X\otimes\one$ to $X\otimes Z$ under conjugation, we find that the stabilizer generators for graph states have the form
\begin{equation}
K_a=X_a \prod_{{b\in{N(a)}}} Z_b
\end{equation}
for every qubit $a$ in the graph. $N(a)$ is the neighbourhood of $a$, i.e. the set of qubits  sharing edges with $a$ on the graph (this corresponds to nearest neighbours in a cluster state).

\subsection{A logical Heisenberg picture}\index{Quantum computation!Heisenberg picture}\index{logical Heisenberg picture}

We are going to use the stabilizer formalism to understand the one-way
patterns which implement unitary transformations in the one-way model. We shall see that it is convenient
to describe logical  action of a
one-way pattern in a \emph{logical  Heisenberg
picture} \cite{brownebriegel_gottheis}. 

The Schr\"{o}dinger picture is the most common approach to describing the time-evolution of quantum systems. Temporal changes in the system are reflected in changes in the state vector or density matrix, e.g. for unitary evolution $\dket{\psi}\mapsto U(t)\dket{\psi}$ or $\rho\mapsto U(t)\dket{\psi}\rho U(t)^\dag$. The observables which characterize measurable quantities, such as Pauli observables $X$, $Y$ and $Z$, remain invariant in time. 
In the Heisenberg picture, on the other hand,  time-evolution is carried exclusively by physical observables which evolve $O(t)\mapsto U(t)^\dag O U(t)$. States and density matrices remain constant in time.

A \emph{logical Heisenberg picture}, also called a ``Heisenberg representation of quantum computation''  \cite{brownebriegel_gottheis}, is a middle-way between these two approaches, containing elements of both. We shall introduce it with an  example, starting in the Schr\"{o}dinger picture with a single-qubit density matrix $\rho(t)$ evolving in time. Since the $n$-qubit Pauli-group operators form a basis in the vector space of $n$-qubit Hermitian operators, we can write $\rho$ at time $t=0$ as
\begin{equation} 
\rho(t=0)=a\,\one+b\,X+c\,Y+d\,Z
\end{equation}
where $a$, $b$, $c$ and $d$ are real parameters which define the state.

At time $t$, the state has been transformed through unitary $U(t)$. In the usual Schr\"{o}dinger picture one would reflect this in a transformation of the matrix elements of the state, or, equivalently, of the parameters $a$, $b$, $c$ and $d$ to  $a(t)$, $b(t)$, etc.. However, one can also write
\begin{equation}
\rho(t)=U(t)\rho U(t)^\dag=a\,\one +b\,  U(t) X U(t)^\dag+c\,  U(t)YU(t)^\dag+d\, U(t)ZU(t)^\dag\ .
\end{equation}

By introducing time-evolving observables $\overline{X}(t)= U(t) X U(t)^\dag$ and similar expressions for $\overline{Y}(t)$ and $\overline{Z}(t)$, we can express this as
\begin{equation} \label{brownebriegel_rhoexpansion}
\rho(t)=a\,\one+b\,\overline{X}(t)+c\,\overline{Y}(t)+d\,\overline{Z}(t)\ .
\end{equation}
The time evolution is thus captured by the evolution of these logical observables, and the parameters $a$, $b$, $c$ and $d$ remain fixed. Since $\overline{X}(t)$,  $\overline{Y}(t)$, etc. define the logical basis in which $\rho$ is expressed, we call them \emph{logical observables}.

Since $\overline{Y}(t)=i\overline{X}(t)\overline{Z}(t)$, determining $\overline{X}(t)$ and $\overline{Z}(t)$ specifies the  evolution  $U(t)$ completely. More generally, an $n$-qubit unitary is defined in this picture by the  evolution of  $\overline{X}(t)_a$ and  $\overline{Z}(t)_a$ for each qubit $a$.  It is important to emphasise that  the logical observables $\overline{X}(t)$, $\overline{Y}(t)$, and so forth, are no longer equal to the physical observables $X$, $Y$ etc. which remain constant in time.
Here time evolution is characterized by the evolution of  \emph{logical} observables. In analogy to the (standard) Heisenberg picture, where  \emph{physical} observables evolve in time, we call this a \emph{logical Heisenberg picture}\footnote{In geometric terms, evolution in the Schr\"{o}dinger picture corresponds to an active transformation of a state. A logical Heisenberg picture corresponds to a passive transformation -- the state remains fixed with respect to a changing  logical basis.}.

The logical Heisenberg picture  can be illustrated with some simple examples. First, let us consider a Hadamard $U(t)=H$. This is represented in the logical Heisenberg picture through $\overline{X}(t)=Z$ and  $\overline{Z}(t)=X$. Second, let us look at the representation of the SWAP gate in this picture. We find that  $\overline{X}_1(t)=X_2$ and  $\overline{X}_2(t)=X_1$ (similarly for the $Z$ variables). The logical Heisenberg picture clearly encapsulates the action of these gates; in the case of the Hadamard, we see  $X$ and $Z$ interchanged and for  SWAP, the operators on the two qubits are switched round. In the one-way quantum computer logical time evolution is discrete and driven by single-qubit measurements, so in the following we will often suppress the  time labelling $t$.

A logical Heisenberg picture becomes particularly useful when describing the  \emph{encoding} of quantum information.  As well as density matrices, one can also represent the evolution of pure state vectors in a logical Heisenberg picture.  The time evolution is carried by the \emph{logical basis} states, the joint eigenstates of $\overline{Z}(t)_a$ with phase relations fixed by $\overline{X}(t)_a$. Consider a state $\dket{\psi}=\alpha\dket{0}+\beta\dket{1}$ imagine we encode it via some unitary transformation $U$. We would  write  $\dket{\psi}=\alpha U\dket{0}+\beta U\dket{1}=\alpha \dket{0'}+\beta \dket{1'}$, where $\dket{0'}$ and $\dket{1'}$ are the new (encoded) logical basis states. Thus ``encoding'' implicitly adopts a logical Heisenberg picture. The state coefficients remain constant while logical basis vectors are transformed.

\subsection{Dynamical variables on a stabilizer sub-space}

This formalism can be combined with the stabilizer formalism to track the evolution of logical observables on a sub-space of a larger system. 
The stabilizer group then defines the logical sub-space, and the dynamical logical operators  track the evolution of this sub-space. The logical operators act only to map states  around the sub-space, therefore they must commute with the stabilizers of that sub-space.

Let us use the well-known three-qubit error correcting code as an example. In this code, the logical $\dket{0}$ is represented by $\dket{0}\dket{0}\dket{0}$ and  $\dket{1}$ by $\dket{1}\dket{1}\dket{1}$. The stabilizer group for this sub-space is generated by $Z\otimes Z\otimes\one$ and  $\one\otimes Z\otimes Z$. The logical observables associated with this basis are $\overline{Z}=Z\otimes\one\otimes\one$  and $\overline{X}=X\otimes X\otimes X$. One can easily verify that these operators have the desired action on the logical basis states. 

However, even though the logical basis is entirely symmetric under interchange of the qubits, the logical $\overline{Z}$ is not. Due to the symmetry of the situation one would expect that $\one\otimes Z\otimes\one$ and   $\one\otimes \one\otimes Z$ would be equivalent to the physical representation of  $\overline{Z}$ we have chosen above. 
That these operators have the same action on the logical basis states  is  easy to confirm and it reflects an important characteristic of logical operators on a sub-space, namely that they are not unique. Given a stabilizer operator for the sub-space $K_a$ and logical operator $\overline{L}$, the product $K_a\overline{L}$ has the same action on the logical sub-space as $\overline{L}$. Thus there are a number of physical representations for a given logical observable. Formally this set is in fact  a coset of the stabilizer group. 
In order to define this set, only one member of the set need be specified. When we write a particular physical operator corresponding to  $\overline{L}$  this is just  a ``representative'' of the whole coset.

\subsection{One-way patterns in the stabilizer formalism}

We introduced the term ``one-way pattern'' to describe a layout of qubits, graph state edges and measurements which implements a given unitary in the one-way model. More specifically, the patterns contain a set of  qubits labelled input qubits and a set labelled output qubits, a set of auxiliary qubits and a set of edges connecting those qubits. 
We will now show how, using the rules for transforming stabilizer sub-spaces under measurement, that the one-way pattern will lead to the transformation of the logical operators $\overline{X}_a\mapsto\pm U\overline{X}_{a'}U^\dag$ and  $\overline{Z}_{a}\mapsto\pm U\overline{Z}_{a'}U^\dag$. This is a logical Heisenberg picture representation of the desired unitary $U$, plus the displacement of the logical state from input qubit(s) $a$ to output qubit(s) $a'$. The extra factor $\pm1$ reflects the presence of by-product operators (due to the randomness of the measurement outcomes) since $XZX=-Z$ and $ZXZ=-X$.

\subsection{Pauli measurements}

\begin{figure}\begin{center}
\includegraphics[scale=0.2]{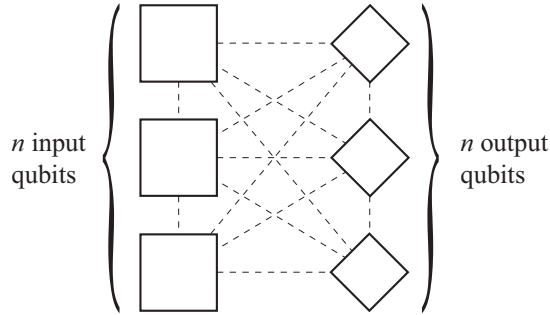}
\end{center}
\caption{\label{brownebriegel_clifford} Any $n$-qubit  Clifford group operation may be implemented (up to local Clifford corrections) by a one-way pattern with $2n$-qubits. Dotted lines represent possible edges in the patterns.}
\end{figure}

Before we consider one-way patterns with general one-qubit measurements, let us first consider patterns consisting solely of Pauli measurements. 
These measurements change the logical variables'  encoding according to the desired evolution of the logical state.
As the logical evolution is unitary, each measurement must reveal no information about the logical state. By considering commutation relations, one can show that these requirements are equivalent to demanding that the measured observable   anti-commute with at least one stabilizer generator.

The effect of performing  a measurement of  a (multi-qubit) Pauli observable $\Sigma$ on a sub-space is as follows (such methods are described in more detail in \cite{brownebriegel_nielsenchuang}). If $\Sigma$ does not commute with the complete stabilizer group, one can always construct a set of stabilizer generators such that only one of the generators $K_a$ anti-commutes with $\Sigma$. 
The stabilizers which commute with $\Sigma$ must also stabilize the transformed sub-space after the measurement, which  will be an eigenspace of $\Sigma$ with eigenvalue $\pm1$. Thus $\pm\Sigma$ will  itself belong to the new stabilizer. We can thus construct a set of generators for the stabilizer of the transformed sub-space, by  simply replacing $K_a$, which anti-commutes with $\Sigma$, with $\pm\Sigma$. 

The logical observables transform in a similar way. This time, just one member of the coset for each logical observable needs to be found which commutes with $\Sigma$. If the representative logical operator $\overline{L}$ commutes with $\Sigma$ it remains a valid representative logical operator after the measurement (the full coset will be different though due to the changed stabilizer). If $\overline{L}$ does not commute with $\Sigma$, then the product  $\overline{L}K_a$ does commute, so logical operators for the transformed sub-space are easy to find.

A final step involves finding a reduced description of the state which ignores the now unimportant measured qubit. This is achieved by choosing a set of stabilizer generators  where all but one ($\pm \Sigma$ itself) act as the identity on the measured qubit. This is achieved by multiplying all the generators not already in this form with $\pm \Sigma$.  In the same way representative logical operators can be chosen that are also restricted to the unmeasured qubits.

After all but the designated output qubits in a pattern have been measured, the one-way pattern has been completed. The reduced description of the output qubits has a stabilizer group consisting of the  identity operator alone and logical operators have become  $\overline{X}_a=\pm U X_{a'}U^\dag$ and  $\overline{Z}_a=\pm UZ_{a'}U^\dag$. We interpret this in the logical Heisenberg picture. The one-way pattern has implemented the unitary transformation $U$ plus known Pauli corrections and the logical sub-space has been physically  displaced  from the input qubits $a$ to output qubits $a'$.

\begin{figure}
\begin{center}
\includegraphics[width=6cm]{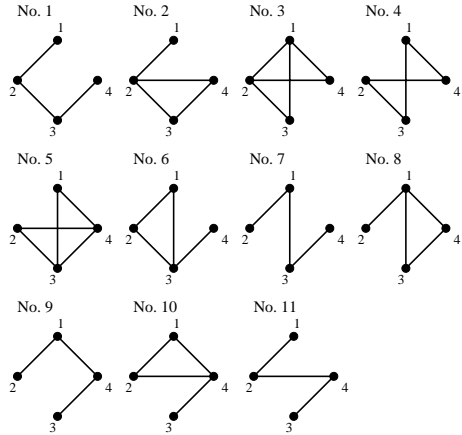}
\end{center}
\caption{\label{brownebriegel_loccomp}The full orbit of locally equivalent four-qubit graph states. Each graph state is obtained from the previous one by application of the ``local complementation rule''.  This figure is taken from M. Hein, J. Eisert and H.J. Briegel, Phys. Rev. A 69, 062311 (2004) \copyright APS.}
\end{figure}

This method can be used to design and verify one-way patterns (e.g. see Exercise 3). It may seem complicated for such simple examples, but its power lies in its generality. In the next section, we show how  measurement patterns for arbitrary Clifford operations may be evaluated using these  techniques.

\subsection{Pauli measurements and the Clifford group}\index{Clifford group}

In the previous section, all of the transformations of the logical observables keep their physical representations within the Pauli group.
Unitary operators which map Pauli group operators to the Pauli group under conjugation are known as Clifford group operations. The Clifford group is the group  generated by the CZ, Hadamard and $U_z(\pi/2)$ gates. Since all of these gates can be implemented by one-way patterns with Pauli measurements only (i.e. by choosing $\phi=0$ or $\phi=\pi/2$ in  figure ~\ref{brownebriegel_simplepatterns} (a)) any Clifford group operation can be achieved by Pauli measurements alone.

The Clifford group plays an important role in quantum computation theory. Clifford group circuits are the basis for most quantum error correction schemes, and many interesting entangled states (including, of course, graph states) can be generated via Clifford group operations alone. However, Gottesman and Knill \cite{brownebriegel_gottheis} showed that notwithstanding this, Clifford group circuits  acting on stabilizer states (such as the standard input $\dket{0}^{\otimes n}$) can be simulated efficiently on a classical computer \cite{brownebriegel_gottesmanaaronson,brownebriegel_anders}. This is  because of the simple way the logical observables transform (in the logical Heisenberg picture) under such operations.

Let us consider the effect of the by-product Pauli operators, generated every time a measurement outcome is $-1$, when Clifford operations are implemented in the one-way quantum computer. Given a Clifford operation $C$, by the definition of the Clifford group, $C\Sigma C^\dag=\Sigma'$ where $\Sigma$ and $\Sigma'$ are Pauli group operators. Therefore $C\Sigma=\Sigma'C$ meaning that interchanging the order of Clifford operators and Pauli corrections will leave the Clifford operation unchanged. This means that there is no need to choose measurement bases adaptively. We thus see that in any one-way quantum computation all Pauli measurements can be made simultaneously in the first measurement round.

These results imply that  Pauli measurements on stabilizer states will always leave behind a stabilizer state on the unmeasured qubits. Additionally,  any stabilizer state can be transformed to a graph state by  local Clifford operations  \cite{brownebriegel_stabgraphequiv,brownebriegel_localequivalence}. Furthermore, this graph state is in general not unique, by further local Clifford operations a whole family of  locally equivalent  graph states can be achieved \cite{brownebriegel_hein,brownebriegel_localequivalence}. The rules for this local equivalence are simple -- a graph can be transformed into another locally equivalent graph by ``local complementation'' \cite{brownebriegel_localequivalence} which is a graph-theoretical primitive \cite{brownebriegel_graphtheory}. In local complementation, a particular vertex of the graph is singled out and the sub-graph given by all vertices connected to it is ``complemented'' (i.e.  all present edges are removed and any missing edges are created). The set of locally equivalent four-qubit graph states is illustrated in figure~\ref{brownebriegel_loccomp}.

This theorem allows us to understand the effect of Pauli measurements on a graph state in a new way. Any Pauli measurement on a graph state simply transforms it  (up to a local Clifford correction) into  another graph state. A graphical description of how the graph is transformed and which local corrections must be applied can  be found in \cite{brownebriegel_hein,brownebriegel_schlingemanngraph}. The rule for $Z$-measurements is simple, the measured qubit and all edges connected to it are removed from the graph. If the -1 eigenvalue was measured, extra $Z$ transformations on the adjacent qubits must be applied to bring the state to graph state form. Rules for $X$ and $Y$ measurements are  more complicated and can be found in \cite{brownebriegel_hein}.

Since the effect of Pauli measurements is to just transform the graph, given  any one-way pattern containing Pauli measurements, the transformation rules can be used to find a one-way pattern which implements the same operation with fewer qubits. The local corrections can often be incorporated in the  bases of remaining measurements. If not they lead to an additional local Clifford transformation on the output qubit.
Since the Pauli measurements correspond to the implementation of Clifford group operations, this leads to a stronger result than the Gottesman-Knill theorem. All Clifford operations, wherever they occur in the quantum computation are reduced to classical pre-processing of the one-way pattern. A further consequence is that any $n$-qubit Clifford group operation can be implemented (up to the local Clifford corrections) on a $2n$ qubit pattern, as illustrated in figure~\ref{brownebriegel_clifford}.

\subsection{Non-Pauli measurements}

The method above does not yet allow us to treat non-Pauli measurements, specified by measurement directions other than along the $X$-, $Y$- or $Z$-axis. However, one can still treat such measurements within the stabilizer formalism. The stabilizer eigenvalue equations (equation~(\ref{brownebriegel_stabeig}))  can be rearranged to generate a family of non-Pauli unitary operations which also stabilize the sub-space \cite{brownebriegel_clusterqc2}. Consider the state $\dket{\psi}$ stabilized by  operator $Z\otimes X$. We rearrange the stabilizer equation as follows
\begin{equation}\begin{split}
Z\otimes X \dket{\psi}&=\dket{\psi}\\
Z\otimes \one \dket{\psi}&=\one\otimes X\dket{\psi}\\
\left(Z\otimes \one - \one\otimes X \right)\dket{\psi}&=0\\
\end{split}
\end{equation}
thus for all $\phi$,
\begin{equation}
\exp\left[i \frac{\phi}{2}(Z\otimes \one - \one\otimes X)\right] \dket{\psi}=\dket{\psi}. 
\end{equation}
Thus we have a unitary $U_z(-\phi)\otimes U_x(\phi)$ which itself stabilizes $\dket{\psi}$. This implies that
\begin{equation}\label{brownebriegel_equnitstab}
U_z(\phi) \otimes\one \dket{\psi}=\one\otimes U_x(\phi)\dket{\psi}. 
\end{equation}

Similar unitaries and similar expressions can be generated from any stabilizer operator. We will show in the next section, how this technique allows a simple analysis of the one-way pattern for general unitaries diagonal in the computational basis, and in fact, the technique allows one to understand any one-way pattern solely within the stabilizer formalism and was used to design and verify many of the gate patterns presented in  \cite{brownebriegel_clusterqc2}.
This indicates that the effect of  non-Pauli measurements in a one-way quantum computation can always be understood as the implementation of a \emph{generalized rotation} $\exp[-i(\phi/2)\Sigma]$ where $\Sigma$ can be any $n$-qubit  Pauli group operator. We shall discuss the consequences of this further in section~\ref{brownebriegel_gatepatternsbeyond}.

\subsection{Diagonal unitaries}\label{brownebriegel_diagon}

 Earlier in the chapter we saw that a CZ gate can be implemented such that the input qubit is also the output qubit. Coinciding input and output qubits in a one-way pattern reduces the size of the pattern so it is natural to ask which unitaries can be implemented this way and one can show (see exercise 4) that it is only those unitaries  diagonal in the computational-basis. In fact, there is a simple one-way pattern for  any diagonal unitary transformation. Any such $n$-qubit operator can be written  (up to a global phase)  in the following form
\begin{equation}\label{brownebriegel_eqdiag}
D_n= \prod_{\vec{m}}\exp[i\frac{\phi_{\vec{m}}}{2}(Z_1)^{m_1} (Z_2)^{m_2}\cdots(Z_n)^{m_n}]
\end{equation}
where $(Z_a)^{m_a}$ is equal to the identity if $m_a=0$ and $Z$ acting on qubit $a$ when  $m_a=1$, and  the sum is over all binary vectors {$\vec{m}$} of length $n$.

\begin{figure}\begin{center}
\includegraphics[scale=0.2]{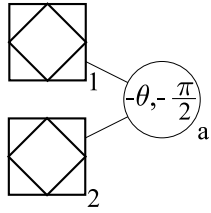}
\end{center}
\caption{\label{brownebriegel_doublez} The one-way pattern which implements the unitary ``double-z rotation''  $e^{-i\frac{\theta}{2}Z\otimes Z}$. Note that input and output qubits coincide.
}
\end{figure}

Each element of this product is a generalized rotation acting on a subset of the qubits and has a very simple implementation in the one-way quantum computer. To illustrate this, consider the two-qubit transformation $e^{-i\frac{\phi}{2}Z\otimes Z}$. 
This can be implemented on a one-way pattern with three qubits as illustrated in figure~\ref{brownebriegel_doublez}. In this pattern the qubits labelled 1 and 2 are the joint input-output qubits, and qubit $a$ is an ancilla. The entanglement graph has two edges connecting $a$ to 1 and 2. A measurement in basis $(-\phi,-\pi/2)$, i.e. of observable $U_x(-\phi)ZU_x(\phi)$, implements $e^{-i\frac{\phi}{2}Z_1 Z_2}$ on the input state, with  by-product operator  $Z_1 Z_2$.

To see this, we recall that the stabilizer for the sub-space corresponding to such a graph is $X_a Z_1 Z_2$. The corresponding eigenvalue equation  can be transformed, as described in the previous section, to generate the stabilizing unitary $[U_x(\theta)]_a e^{i\frac{\theta}{2}Z_1 Z_2}$. Measuring qubit $a$ in basis  $(-\phi,-\pi/2)$ is equivalent to performing $[U_x(\phi)]_a$ and then measuring $Z_a$, hence the one-way pattern implements the logical unitary  $e^{-i\frac{\phi}{2}Z_1 Z_2}$.

\begin{figure}\begin{center}
\includegraphics[scale=0.2]{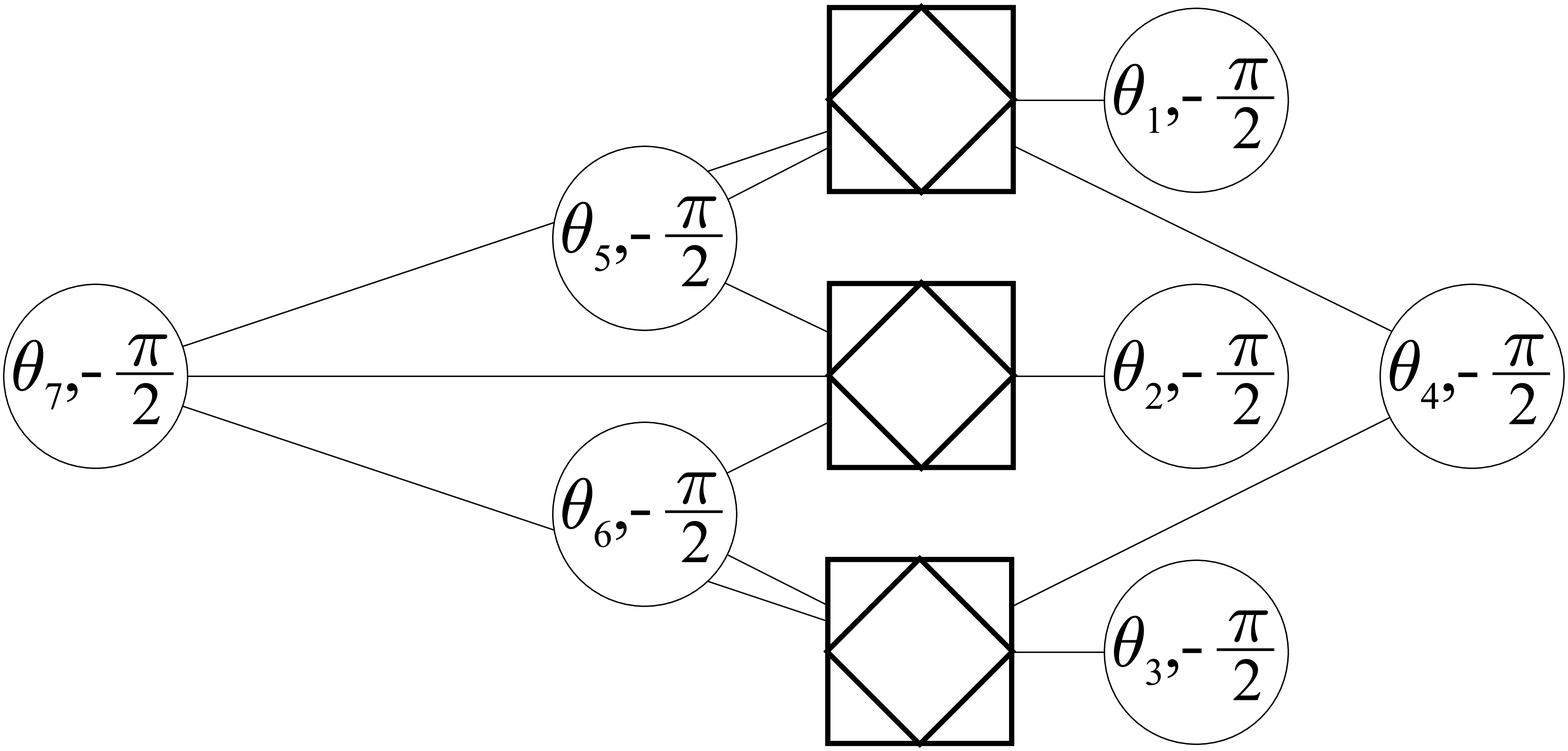}
\end{center}
\caption{\label{brownebriegel_diagonal} Arbitrary diagonal unitaries, may be implemented in a single  round of measurements by measurement patterns with coinciding input and output qubits. This example shows an arbitrary diagonal three-qubit unitary $\exp[\tfrac{i}{2}({\theta_1}Z\otimes\one\otimes\one+{\theta_2}\one\otimes Z\otimes\one+{\theta_3}\one\otimes\one\otimes Z +{\theta_4}Z\otimes\one\otimes Z +{\theta_5}Z\otimes Z\otimes\one +{\theta_6}\one\otimes Z\otimes Z+ {\theta_7}Z\otimes Z\otimes Z)]$. For example, by setting the angles to $\theta_1=\theta_2=\theta_3=\theta_7=-\pi/4$ and $\theta_4=\theta_5=\theta_6=\pi/4$, we obtain a control-control Z gate or ``Toffoli-Z gate''. See \cite{brownebriegel_clusterqc2} for a cluster-state implementation of this gate.}
\end{figure}

We can generalize this pattern to quite general $n$-qubit diagonal unitaries. (Verify this in exercise 5). For example, the pattern for an arbitrary diagonal three-qubit unitary is given in figure~\ref{brownebriegel_diagonal}.  This is  a highly parallelized  and efficient implementation of the unitary\footnote{This is reminiscent of the results reported in \cite{brownebriegel_chrismoor} regarding the parallelization of diagonal unitaries, where, however  a different definition of parallelization is used. We treat the CZ operations generating the graphs state as occurring in a single time-step. Physically this is entirely reasonable as operations generated by commuting Hamiltonians can often be implemented simultaneously as we shall see in our discussion of optical lattices in section~\ref{brownebriegel_implementations}.}. 

Since the by-product operators for these patterns are diagonal themselves  they  commute with the desired logical diagonal unitaries.  Thus there is no dependency in the measurement bases on the outcome of measurements within this pattern and all measurements can be achieved in a single measurement round. Thus, not only can a quantum circuit consisting of Clifford gates alone be implemented in a single-time step -- this is true for any diagonal unitary followed by a Clifford network.

\subsection{Gate patterns beyond the standard network model -- CD-decomposition}\label{brownebriegel_gatepatternsbeyond}\index{CD-decomposition of unitary operators}

We have seen that one can construct one-way patterns to implement a unitary operation described by a quantum circuit by simply connecting together patterns for the constituent gates. Furthermore, such patterns can be made more compact by evaluating the graph state transformations corresponding to any Pauli measurements present. This can change the structure of the pattern such that the original circuit is hard to recognize (see for example, the quantum Fourier transform patterns in \cite{brownebriegel_hein}).

We have also seen that non-Pauli measurements in a measurement pattern lead to generalized rotations on the logical state of the form $\exp[-(i/2)\phi\Sigma]$ where $\Sigma$ is some Pauli group operator. The implementation of any non-Clifford unitary on the one-way quantum computer is thus best understood as a sequence of operators of this form. Two such operators $\exp[-(i/2)\phi\Sigma]$ and  $\exp[-(i/2)\phi'\Sigma']$ may, if $[\Sigma,\Sigma']=0$ be combined to give  $\exp[-(i/2)[\phi\Sigma+\phi'\Sigma']$. In general, any operators of the form  $\exp[(i/2)[\sum_a\alpha_a\Sigma_a]$, where $[\Sigma_a,\Sigma_{a'}]=0$, can be diagonalized by a Clifford group element $C$ to $CDC^\dag$, where $D$ is a diagonal unitary. Composing two operations this form, e.g  $C_1D_1C_1^\dag$ and  $C_2D_2C_2^\dag$ will give  $C_1D_1C_1^\dag C_2D_2C_2^\dag=C_1D_1C_3D_2C_2^\dag$ where $C_3=C_1^\dag C_2$, and we call the casting of a unitary in this form  a \emph{CD-decomposition}.

There are several observations to be made about such decompositions. We have already seen that both diagonal unitaries and Clifford group operations have compact implementations in one-way quantum computations. This means that CD-decompositions are very useful in the design of compact one-way patterns. One simply combines te one-way patterns for diagonal unitaries presented above with patterns for Clifford operations, which we have seen require at most $2n$ qubits for an $n$-qubit operation and which can be constructed either by employing Pauli transformation rules on a pattern for a network of  CZ, Hadamard and $U_z(\pi/2)$ gates, or by inspection of the logical Heisenberg form of the operation. 
%
In \cite{brownebriegel_clusterqc2} this decomposition, together with stabilizer techniques described above, was used to design cluster-state implementations for several gates and simple algorithms including controlled Z-rotations and the quantum Fourier transform (QFT).
%

 A further advantage in working with a CD-decomposition is that it immediately provides an upper bound in the number of time steps needed for the implementation of the one-way pattern. This is simply the number of ``CD units'' in the decomposition. We saw in section~\ref{brownebriegel_diagon} that a single CD unit can be implemented in a single time step. Each CD unit in turn will create by-product operators, which may need to be accounted for in the choice of measurement bases for following  diagonal unitaries.  A decomposition which minimised the number of CD units would give a (possibly tight) upper bound on the minimal number of time-steps and would be one measure of how hard the unitary is to implement in the one-way model. For example, Euler's rotation theorem tells us that the optimal CD-decomposition for an arbitrary rotation consists of three CD units and correspondingly requires three measurement rounds for implementation on the one-way quantum computer.

Note that there is considerable freedom in choosing a CD form. For example, one can construct the decomposition such that all the diagonal gates are solely local, single-qubit operations and only the Clifford gates are non-local. This gives a degree of flexibility in the design one-way patterns.

Quantum circuits described in terms of  Clifford group gates plus rotations can readily cast in CD form by decomposing the rotations into $Z$-axis rotations and Hadamards. One can then reduce the size of the corresponding pattern by applying the  Pauli measurement transformation rules\footnote{It is important to note, however, that the Pauli measurement rules alone do not usually provide a CD-decomposition which is optimal in the sense of consisting of the smallest number of CD units. An optimal CD-decomposition will allow the construction of a one-way pattern for the unitaries with fewer measurement rounds, and often a more compact entanglement graph, than  application of the Pauli transformation rules alone.}. Quantum circuits for the simulation of general Hamiltonians are usually expressed using the Trotter formula (see \cite{brownebriegel_nielsenchuang}) which leads to unitaries which are a sequence of generalized rotations which can be cast in CD form in a straightforward manner. Thus the one-way quantum computer is very well suited to Hamiltonian simulation (see e.g. \cite{brownebriegel_durbremner}), which will be an important application of quantum computers.

\section{Implementations}\label{brownebriegel_implementations}
\subsection{Optical lattices}\index{Optical lattice}

Beyond its theoretical  value, there are a number of physical implementations where one-way quantum computation gives distinct practical advantages. One of these is in systems where graph states or cluster states can be generated efficiently, such as ``optical lattices''. In an optical lattice, cold neutral atoms are trapped in a lattice structure, given by the periodic potential due to a set of superposed laser fields. The potential ``seen'' by each atom depends on its internal state. This means that  neighbouring atoms in different states can be brought close together by changing the relative positions of the minima of the periodic potentials, leaving an interaction phase on the atoms' state \cite{brownebriegel_dieterhanscollision}. If this is timed such that this interaction phase is $-1$ the process implements, essentially, a CZ gate between the two atoms. However, every atom in the lattice will be affected when  these potentials move and thus CZ gates can be implemented between neighbouring qubits across the lattice simultaneously. Thus, by preparing all atoms in a superposition of these internal states beforehand, a very large cluster state can  be generated very efficiently. In recent years there has been much progress in the generation and manipulation of ultra-cold atoms in optical lattices in the laboratory \cite{brownebriegel_optlatexp}, and a number of schemes for the generation of arbitrary graph states in these systems have been proposed \cite{brownebriegel_optlattprop}. Possibly the most difficult obstacle to overcome for the implementation of one-way quantum computation in optical lattices is the difficulty in addressing individual atoms in the lattice.

\subsection{Linear optics and cavity QED}\index{Quantum computation! Linear optical}\index{Linear optical quantum computation}\index{Cavity QED}

Photons make excellent carriers on quantum information and are relatively decoherence free. A key difficulty in implementing universal quantum computation using photons is that two-qubit gates such as CZ cannot be implemented by the simple linear optical elements of the optics laboratory (e.g. beam-splitters and phase shifters) alone. By employing photon number measurements, non-deterministic entangling gates are possible. Most times, however,  the gate fails, and this failure leads to the  measurement of the qubits' state which disrupts the computation. Naively, one would expect that scaling this up into a circuit would lead to an exponential decrease in success probability, but, by using a combination of techniques including gate teleportation \cite{brownebriegel_gottesmanchuang} and error correction, scalable quantum computation is indeed possible \cite{brownebriegel_klm}. A key disadvantage of this particular approach, however,  is that each gate requires a large number of ancilla photons in a difficult-to-prepare entangled state.

A much more efficient strategy is to use the non-deterministic gates to build an entangled resource state for measurement-based quantum computation   \cite{brownebriegel_yoranreznik,brownebriegel_nielsencluster}. Cluster states can be generated efficiently  \cite{brownebriegel_efficientTezDan} using  so-called ``fusion operations'',\index{Fusion operations on graph states} which can be performed (non-deterministically) with simple  linear optics. Fusion operations \cite{brownebriegel_efficientTezDan,brownebriegel_verstraete} are implementations of operators such as $\dket{0}\dbra{00}+\dket{1}\dbra{11}$, which when applied to two qubits in different graph states,  replace both qubits  by a single one which inherits all the graph state edges of each, thus ``fusing'' the two graph states together. Recently, three and four-qubit graph states have been created in the laboratory  using methods based on down-conversion and post-selection \cite{brownebriegel_zeilinger} and fusion measurements \cite{brownebriegel_zhang}. Single-qubit measurements on these states demonstrated many of the key elements of one-way quantum computation \cite{brownebriegel_zeilinger}. More details of linear optical quantum computation can be found in other chapters of this book and in a recent review \cite{brownebriegel_kokreview}.

Quantum computation with photons is not the only scenario where gates are inherently non-deterministic. Similar techniques can be used to implement non-deterministic gates between atoms or ions trapped in separate cavities. Cavity QED implementations of the one-way quantum computer is a fast-developing area and recently there have been a number of promising experimental proposals \cite{brownebriegel_cavityqed}.

\section{Recent developments}\label{recentdev}

In addition to these developments toward the implementation of one-way quantum computation there have been a number of interesting papers exploring its theoretical structure. The relationship between the one-way quantum computer and other models of measurement-based quantum computation \cite{brownebriegel_nielsenmeas,brownebriegel_gottesmanchuang} has been explored in \cite{brownebriegel_perdrix,brownebriegel_panos,brownebriegel_childs,brownebriegel_verstraete} and  an algebraic representation of one-way graph measurement patterns \cite{brownebriegel_kashefidanos} has been developed.   The simulability of one-way quantum computations with one-way patterns of various depths and geometries has been investigated \cite{brownebriegel_sim,brownebriegel_nielsenreview}.
Looking beyond qubit implementations, an analogue of graph states in continuous-variable harmonic oscillator systems \cite{brownebriegel_zhangcont} has been investigated and  generalizations of one-way quantum computation to $d$-level systems \cite{brownebriegel_qudit} have been explored. A version of one-way quantum computation based on three-qubit interactions has been proposed \cite{brownebriegel_tame}.

Any practical quantum computation proposal must be able to function in the presence of a degree of experimental noise and decoherence. Standard approaches to fault-tolerant quantum computation have been firmly rooted in the network model and it was not clear whether they would translate to the one-way model.  It was shown in \cite{brownebriegel_rausthesis} and later \cite{brownebriegel_dawson} that
 physical errors  in the one-way quantum computer would be manifested as logical errors quite different from those that one would expect in a standard gate network implementation.
Nevertheless for a number of physically reasonable independent noise models there is an error threshold below which fault-tolerant quantum computation is possible \cite{brownebriegel_rausthesis,brownebriegel_dawson}. A simple proof of this for both Markovian and non-Markovian local errors is presented in \cite{brownebriegel_leungaliferis}. 
The implementation of these techniques in linear optical quantum computation has been simulated \cite{brownebriegel_dawsonhaselgrove}, leading to estimated error thresholds of around $0.0001$ for depolarisation errors and $0.003$ for loss. 

Recently, a different  approach to fault-tolerance in the one-way model has been taken. Most quantum error correcting code-words are stabilizer states, and as we have seen, every stabilizer state is locally equivalent to a graph state. It is therefore natural to look for error correction schemes which make use of the natural error correcting properties of the graph state. It has been demonstrated that one-way quantum computation  with a high degree of robustness against qubit loss errors  (the most significant error source in linear optical quantum computation) can be acheived by using a graph states with a tree-like structure  \cite{brownebriegel_varnava}. This scheme tolerates losing up to half of the qubits in the graph state, and can be applied to deal with photon loss errors in linear optical proposals \cite{brownebriegel_varnavaprep}.

Most recently, it was shown  \cite{brownebriegel_raussenfault} that a three-dimensional body-centred cubic lattice cluster state has the properties of a topological surface code. By combining ideas from topological quantum computation with the observation that quantum Reed-Muller codes \cite{brownebriegel_bravyi} allow fault-tolerant non-Pauli measurements of logical qubits to be implemented by local measurements, a fully fault-tolerant scheme was presented \cite{brownebriegel_raussenfault} with estimated error thresholds between 0.001 and 0.01. 

\section{Outlook}

In this chapter, we have given an introduction to the key ideas of one-way quantum computation and some of the most useful mathematical techniques for describing and understanding it. The one-way approach has provided a new paradigm for quantum computation which is casting many questions of quantum computation theory in a new  light. It is leading to experimental implementations that are radically different from early ideas about how a quantum computer would operate. In addition, it is likely that there will  be  further physical systems in which the one-way model offers the most achievable path to quantum computation.  Not least, 
the success of the one-way approach illustrates the power of novel representations of quantum information processing and 
should encourage us to look for other new and distinct models of quantum computation.

\section*{Acknowledgements}

Dan Browne is supported by Merton College, Oxford and EPSRC's QIPIRC programme.
Hans Briegel is supported by the Austrian Science Fund (FWF), the German
Science Foundation (DFG), and by the European Union through projects QLAQUI
and SCALA.

We would  like to thank Robert Raussendorf for many insightful discussions over a number of years, which have helped to shape our  perspective of one-way quantum computation.
Dan would like to thank Sean Barrett, Simon Benjamin, Michael Bremner, Jens Eisert, Joe Fitzsimons,  Elham Kashefi,  Pieter Kok, Michael Nielsen, Terry Rudolph and Michael Varnava for some illuminating discussions about one-way quantum computation. 
Hans would like to thank the Kavli Institute for Theoretical Physics (KITP)
for their hospitality and support while the final part of this work was
completed.  
We  would  like to thank Earl Campbell for helpful comments on the manuscript.

\section*{Exercises}
\begin{enumerate}
\item There are only two topologically distinct three-qubit graph states. In one, the qubits form a linear three qubit cluster state, in the other, the qubits are connected in a triangle. Write down the stabilizer generators for these two states and hence also the full stabilizer group for each. Now show that one can transform between these two states by a local Clifford operator.

\item  Prove that, to generate the stabilizer group for  a $k$-qubit stabilizer sub-space in an $n$ qubit system, $(n-k)$ generators are required.

\item Consider the one-way pattern illustrated in figure~\ref{brownebriegel_simplepatterns}(a) with angle $\phi$ set to zero. Show that after the entangling CZ operation,  but before the measurement, the logical operators $\overline{X}$ and  $\overline{Z}$ have physical representations $X\otimes Z$ and $Z\otimes\one$ respectively. Find the stabilizer and hence the full coset of each logical observable. When observable $X$ is measured on the first qubit, how are the stabilizer and logical observables transformed? Hence verify that this pattern implements a Hadamard gate.

\item Show that one-way  patterns where all input and output qubits coincide can only implement diagonal unitaries. What can one say about patterns where only some of the input and output qubits coincide?

\item Using the decomposition of an arbitrary $n$-qubit diagonal unitary $D_n$ in equation~(\ref{brownebriegel_eqdiag}) and by generalising the methods in section~\ref{brownebriegel_diagon} describe  a one-way pattern which implements $D_n$  requiring a total of $n+(2^n-1)$ qubits.


\item Verify the effect of  applying the ``fusion'' operator $|0\rangle\langle0|\langle0|+|1\rangle\langle1|\langle1|$ to two qubits, each of which belong to seperate graph states. What happens when a projection onto the even-parity sub-space $|0\rangle|0\rangle\langle0|\langle0|+|1\rangle|1\rangle\langle1|\langle1|$  is applied instead?

\item Consider a qubit that is prepared in an unknown state, and a one-dimensional cluster state. What is the effect of applying a fusion operator on the unknown qubit and the qubit at one end of the cluster state. How can the fusion operator be used to ``input''
 externally provided states into a one-way quantum computation?

\end{enumerate}




\end{document}